\documentclass[twocolumn]{aastex7}

\usepackage{bm}
\usepackage{amsmath}
\usepackage{amssymb}
\DeclareMathOperator{\sech}{sech}
\usepackage{float}

\begin{document}

\title{Unveiling a New $\beta$ Scaling of the Tearing Instability in Weakly Collisional Plasmas}

\author[orcid=0000-0001-7144-5777,sname='Ferreira-Santos']{Gabriel L. Ferreira-Santos}
\affiliation{Astrophysics Division, National Institute for Space Research, São José dos Campos, SP, Brazil}
\email{gabriel.ferreira@inpe.br}

\author[orcid=0000-0002-0176-9909,sname='Kowal']{Grzegorz Kowal}
\affiliation{Escola de Artes, Ciências e Humanidades, University of São Paulo, São Paulo, SP, Brazil}
\email{grzegorz.kowal@usp.br}

\author[orcid=0000-0002-1914-6654,sname='Falceta-Gonçalves']{Diego A. Falceta-Gonçalves}
\affiliation{Escola de Artes, Ciências e Humanidades, University of São Paulo, São Paulo, SP, Brazil}
\email{dfalcetal@usp.br}

\correspondingauthor{Gabriel L. Ferreira-Santos}

\begin{abstract}
We investigate the linear tearing instability in weakly collisional plasmas using a nonideal gyrotropic MHD framework, uncovering a previously unknown scaling relation for the instability growth rate in high-$\beta$ environments. Even starting from an isotropic equilibrium, our analysis reveals a $\beta$-dependence, with the maximum growth rate scaling as $\sigma_\mathrm{max} \tau_a \propto \beta^{-1/4}$, challenging the long-held assumption of $\beta$-independence inherent in classical MHD formulations. This novel scaling emerges due to self-consistent fluctuations in pressure anisotropy, dynamically induced by perturbations in velocity and magnetic fields. Increasing plasma-$\beta$ always suppresses the instability, whereas a background pressure anisotropy can either enhance or further suppress it, depending on its sign: for $p_{\parallel,0} < p_{\perp,0}$ the instability is strengthened, while for $p_{\parallel,0} > p_{\perp,0}$ it is weakened. Importantly, this effect is not limited to low-collisionality plasmas at high $\beta$; it can also manifest in more collisional environments once the strict assumption of pressure isotropy is relaxed. This finding has profound implications for various astrophysical contexts characterized by high $\beta$ and varying degrees of collisionality, including the solar corona and heliospheric current sheets, planetary magnetospheres as probed by space missions, and the intracluster medium, where magnetic reconnection critically impacts magnetic field evolution and cosmic-ray transport. Our results therefore question the universality of the widely accepted plasmoid-mediated fast reconnection paradigm and underscore the necessity of incorporating pressure anisotropy effects into reconnection models for accurate representation of astrophysical plasmas.

\end{abstract}



\section{Introduction} 

The tearing mode instability, a resistive magnetohydrodynamic (MHD) process first analyzed by \cite{fkr1963}, leads to the destabilization of a planar current sheet at the vanishing loci of antiparallel components of a magnetic field and the formation of magnetic islands known as plasmoids. Their analysis demonstrated that the growth rate scales as $\sigma \propto S^{-3/5}$ in the short-wavelength regime and as $\sigma \propto S^{-1/3}$ in the long-wavelength regime, where $S = a V_{\rm A} / \eta$ is the Lundquist number, with $V_{\rm A}$ being the characteristic Alfvén speed, $a$ the thickness of the current sheet, and $\eta$ the magnetic resistivity. Furthermore, \cite{fkr1963} demonstrated that the maximum growth rate satisfies $\sigma_\mathrm{max} \tau_a \sim S^{-1/2}$ and $k_\mathrm{max} a \sim S^{-1/4}$, where $\tau_a = a / V_{\rm A}$ is the Alfvén time. Within the incompressible MHD framework, the maximum growth rate and wavenumber of the tearing instability are independent of the plasma-$\beta$, which is defined as the ratio of thermal pressure to magnetic pressure ($\beta = p / p_{\rm mag}$).

Since the pioneering work of \cite{fkr1963}, the linear tearing mode instability has been extensively studied in planar geometry, incorporating various physical effects relevant to laboratory and astrophysical plasmas. In the MHD regime, the inclusion of 3D geometry and a guide field introduces multiple resonant surfaces within the current sheet, enabling oblique mode growth absent in 2D configurations, while a strong guide field can suppress normal variations in the eigenfunctions and modify wave propagation, particularly in the presence of Hall effects \citep{2012PhPl...19b2101B, 2017ApJ...845...25P, Shi2020}.

As mentioned above, these previous studies were performed in the domain of the MHD approximation, where pressure isotropy is assumed presuming a high-collisionality condition. However, observational data from space plasmas confirm their predominantly collisionless nature. Missions such as ESA's Cluster \citep{Escoubet2001} and NASA's THEMIS and MMS \citep{Angelopoulos2008, Burch2016} have resolved kinetic-scale processes, including magnetic reconnection and turbulence, while the Van Allen Probes \citep{Mauk2013} revealed collisionless wave--particle interactions in the radiation belts. Observations from the Parker Solar Probe further consolidated this paradigm, demonstrating the prevalence of magnetically oriented anisotropy in the solar wind pressure tensor \citep{2020ApJS..246...70H}.

It is well established that physical processes occurring at kinetic scales, such as the ion skin depth ($d_i$), are fundamental to understanding fast magnetic reconnection. Effects encapsulated in extended fluid models, most notably the Hall MHD, can become critical. For instance, in the nonlinear evolution of tearing modes, the formation of thin current sheets can be significantly accelerated by Hall physics, leading to a more explosive dynamic on a macroscopic scale \citep{2019ApJ...885...56P}. The focus of the present work, however, is on the linear phase of the tearing instability at its maximum growth rates, which are typically found at length scales much larger than the kinetic scales (i.e., $a \gg d_i$). Within this large-scale approximation, where two-fluid effects have not yet come to dominate the initial growth phase, a single-fluid description---such as nonideal gyrotropic MHD---remains a valid and appropriate framework. This approach allows us to effectively isolate and analyze the role of thermodynamic effects, namely, pressure anisotropy, on the onset of the tearing instability.

Pressure anisotropy, different from the small-scale phenomena mentioned above, may be relevant for the large-scale evolution of tearing in astrophysical scenarios. This work addresses the critical question: how does tearing instability evolve in gyrotropic plasmas? We systematically quantify the effects of anisotropy on growth rates. Our primary objective is to determine whether classical parameter dependencies -- on plasma-$\beta$, the Lundquist number $S$, the anisotropy degree $\Delta\beta = \beta_\parallel - \beta_\perp$, and the magnetic Prandtl number $Pr_{\rm m}$ -- remain unchanged or vary in this gyrotropic regime. Our approach is to first derive the relevant dependencies analytically, which are then validated through numerical simulations.

This paper is organized as follows. Section 2 introduces the nonideal gyrotropic MHD model and derives the full set of linearized equations governing perturbations. In Section 3, we develop an analytical theory of the tearing instability within this framework, applying a boundary-layer analysis to obtain new scaling laws for both the maximum growth rate and the most unstable wavenumber. Section 4 validates these predictions through numerical solutions of the complete eigenvalue problem, providing a systematic assessment of the roles played by plasma-$\beta$, equilibrium pressure anisotropy, and the Lundquist number. In Section 5, we interpret the physical mechanisms underlying the high-$\beta$ stabilization and discuss their broader astrophysical implications. Finally, Section 6 summarizes the principal conclusions of this work.

\section{Nonideal Gyrotropic MHD Model and Linearized Equations}
\label{sec:equations}

We present the governing equations of the nonideal gyrotropic MHD model, which serve as the basis for analyzing tearing instability in magnetized plasmas with gyrotropic pressure. Assuming the incompressible limit ($\nabla \cdot \mathbf{v} = 0$), these equations take the following form:
\begin{equation}
\frac{\partial \mathbf{v}}{\partial t} +(\mathbf{v} \cdot \nabla ) \mathbf{v} = - \nabla \cdot \left( p_\perp \mathbf{I} + \Delta p \mathbf{b} \mathbf{b} \right) + \mathbf{J} \times \mathbf{B} + \nu \nabla^2 \mathbf{v}
\label{eq:momentum}
\end{equation}
\begin{equation}
\frac{\partial \mathbf{B}}{\partial t} =\nabla\times(\mathbf{v}\times\mathbf{B}) +\eta \nabla ^{2}\mathbf{B}, \quad \nabla \cdot \mathbf{B} = 0
\label{eq:induction}
\end{equation}
where $\mathbf{v}$ and $\mathbf{B}$ represent the velocity and magnetic field, respectively; $\mathbf{J} = \nabla \times \mathbf{B}$ is the current density; $\Delta p = p_\parallel - p_\perp$, where $p_\parallel$ and $p_\perp$ are the pressure components parallel and perpendicular to the magnetic field, respectively; $\mathbf{b} \equiv \mathbf{B}/ \left| \mathbf{B} \right|$; and $\nu$ and $\eta$ denote viscosity and magnetic resistivity, respectively.

The nonideal gyrotropic MHD equations are closed by incorporating the evolution equations for the two pressure tensor components, as described in \cite{hau2002}:
\begin{align}
\frac{\partial p_\parallel}{\partial t} & + \left( \mathbf{v} \cdot \nabla \right) p_\parallel = - \left( \gamma_\parallel - 1 \right) p_\parallel \left[ \mathbf{b} \cdot \left( \mathbf{b} \cdot \nabla \mathbf{v} \right) \right] \label{eq:perp_pressure} \\
& + \eta \left( \gamma_\parallel - 1 \right) \left( \mathbf{b} \cdot \mathbf{J} \right)^2 + \frac{1}{3} \nu \left( \gamma_\parallel - 1 \right) \left( \nabla \times \mathbf{v} \right)^2 \nonumber \\
\frac{\partial p_\perp}{\partial t} & + \left( \mathbf{v} \cdot \nabla \right) p_\perp = \left( \gamma_\perp - 1 \right) p_\perp \left[ \mathbf{b} \cdot \left( \mathbf{b} \cdot \nabla \mathbf{v} \right) \right]  \label{eq:parallel_pressure} \\
& + \eta \left( \gamma_\perp - 1 \right) \left[ \mathbf{J} \cdot \mathbf{J} - \left( \mathbf{b} \cdot \mathbf{J} \right)^2 \right] + \dfrac{2}{3} \nu \left( \gamma_\perp - 1 \right) \left( \nabla \times \mathbf{v} \right)^2 \nonumber
\end{align}
where $\gamma_\parallel$ and $\gamma_\perp$ are the adiabatic indices along and across the magnetic field, respectively.

The classical double-adiabatic formulation of \cite{Chew1956}, commonly referred to as CGL-MHD, is known to be inconsistent with observations. To address this limitation, subsequent models have incorporated dissipative effects, such as the extended gyrotropic plasma model by \cite{hau2002}. Following this approach, our work employs more appropriate equations for the parallel and perpendicular pressures that include both ohmic and viscous heating. This method provides a self-consistent framework that conserves total energy by explicitly tracking its transfer from magnetic and kinetic reservoirs into thermal energy.

In this work, we assume a 2.5D geometry (XZ-plane), where all $y$-derivatives are zero. The current sheet is defined by an equilibrium magnetic field $\mathbf{B}_0 = B_0(z)\hat{\mathbf{i}} + B_g(z)\hat{\mathbf{j}}$, where $B_0(z) = \tanh(z/a)$ and $B_g(z) = \sech(z/a)$, ensuring that $\left| \mathbf{B}_0 \right|^2 = 1$. The parameter $a$ represents the current sheet thickness and is set to unity in our analysis. The equilibrium profiles for parallel and perpendicular pressure are uniform. Specifically, the parallel pressure is given by $p_{\parallel,0} = \frac{1}{2} \beta_0 + \frac{1}{2} \Delta \beta_0$, while the perpendicular pressure is $p_{\perp,0} = \frac{1}{2} \beta_0$, where $\beta_0$ and $\Delta \beta_0$ are free parameters. We should note that, in order to ensure $p_\parallel > 0$, we impose the condition $\Delta \beta_0 > -\beta_0$. Initially, equilibrium velocity is zero everywhere.

All field components are separated into the static equilibrium field, and small perturbations of the form $\delta f(t,\mathbf{r}) = \delta f(z) \exp[i k x + \sigma t]$ are assumed, where $k$ and $\sigma$ denote the wavevector's $x$-component and the growth rate of a given mode, respectively. Therefore, the linearization of Eqs.~(\ref{eq:momentum})-(\ref{eq:parallel_pressure}) yields five coupled equations governing the perturbations of velocity, magnetic field, and pressure anisotropy $\Delta p$:

\begin{widetext}
\begin{align}
\label{dvy}
& \sigma \delta v_y = \underbrace{i k B_0 \delta B_y + B_g' \delta B_z}_{\text{ideal MHD part}} \underbrace{- i k B_0 B_g \delta \Delta p}_{\text{pressure anisotropy fluctuations}} \underbrace{- \frac{\Delta \beta_0}{2} \left[ i k B_0 \left( 1 - 2 B_g^2 \right) \delta B_y + B_g' \delta B_z + 2 B_0^2 B_g \delta B_z' \right]}_{\text{equilibrium pressure anisotropy}} \\
& \hspace{1.2in} \underbrace{+ \nu \left( \delta v_y'' - k^2 \delta v_y \right)}_{\text{viscous part}} \nonumber
\end{align}
\begin{align}
\label{dvz}
& \sigma \left( \delta v_z'' - k^2 \delta v_z \right) =
 \underbrace{i k \left[ B_0 \left( \delta B_z'' - k^2 \delta B_z \right) - B_0'' \delta B_z \right]}_{\text{ideal MHD part}} \underbrace{- k^2 B_0 \left[ 2 B_0' \delta \Delta p + B_0 \delta \Delta p' \right]}_{\text{pressure anisotropy fluctuations}} \\
& \hspace{0.8in} \underbrace{- \frac{\Delta \beta_0}{2} \Bigl\{ i k \left[ B_0 \left( \delta B_z'' - k^2 \delta B_z \right) - B_0'' \delta B_z \right] - 2 \left[ i k B_0^2 \left[ B_0 \delta B_z'' + 3 B_0' \delta B_z' \right] \right.}_{\text{equilibrium pressure anisotropy}} \nonumber \\
& \hspace{0.4in} \quad \underbrace{\left. + k^2 B_0^2 B_g \delta B_y' + k^2 B_0 \left[ B_0 B_g' +2 B_0' B_g \right] \delta B_y \right] \Bigr\}}_{\text{equilibrium pressure anisotropy}} \underbrace{+ \nu \left( \delta v_{z}'''' - 2 k^2 \delta v_{z}'' + k^4 \delta v_{z} \right)}_{\text{viscous part}} \nonumber
\end{align}
\begin{align}
\label{dby}
& \sigma \delta B_y = \underbrace{i k B_0 \delta v_y - B_g' \delta v_z}_{\text{ideal MHD part}} \underbrace{+ \eta \left( \delta B_y'' - k^2 \delta B_y \right)}_{\text{resistive part}}
\end{align}
\begin{align}
\label{dbz}
& \sigma \delta B_z = \underbrace{i k B_0 \delta v_z}_{\text{ideal MHD part}} \underbrace{+ \eta \left( \delta B_z'' - k^2 \delta B_z \right)}_{\text{resistive part}}
\end{align}
\begin{align}
\label{dp_lin}
& \sigma k \delta \Delta p = \underbrace{\frac{1}{2} \left( \gamma_\parallel + \gamma_\perp - 2 \right) \beta_0 k B_0 \left( B_0 \delta v_{z}' - i k B_{g} \delta v_{y} \right)}_{\text{plasma-$\beta$ dependence}} \underbrace{+ \frac{1}{2} \left( \gamma_\parallel - 1 \right) \Delta \beta_0 k B_0 \left( B_0 \delta v_{z}' - i k B_{g} \delta v_{y} \right)}_{\text{equilibrium pressure anisotropy}} \\
& \hspace{0.1in} \underbrace{- 2 \eta \left\{ 3 \left( B_g B_0' - B_0 B_g' \right) \left\{ k \left( B_{0}' \delta B_{y} - B_0 \delta B_{y}' \right) \right. + i \left[ B_g \left( \delta B_{z}'' - k^2 \delta B_{z} \right) - B_{g}' \delta B_{z}' \right] \right.}_{\text{ohmic heating}} \nonumber \\
&  \hspace{0.24in} \underbrace{\left. - \left( B_g B_0' - B_0 B_g' \right) \left( i B_0 \delta B_{z}' + k B_g \delta B_{y} \right) \right\} \left. + \left[ k B_{g}' \delta B_{y}' + i B_{0}' \left( \delta B_{z}'' - k^{2} \delta B_{z} \right) \right] \right\}}_{\text{ohmic heating}} \nonumber
\end{align}
\end{widetext}

In the above equations, $'$, $''$, and $''''$ denote the first-, second-, and fourth-order derivatives with respect to $z$, respectively. The details of this derivation are presented in the Appendix.

\section{Analytical Theory of the Tearing Instability in Gyrotropic MHD}
\label{sec:theory}

The stability of a 1D equilibrium current sheet to 2D tearing modes is investigated through a boundary-layer analysis \citep[e.g.][]{fkr1963, Somov2013, 2018JPhCS1100a2003B}. In this approach, the plasma domain is separated into two layers in order to isolate the distinct physical processes that govern the instability. The outer layer spans most of the plasma, where resistivity is negligible and the perturbations evolve according to ideal MHD, but the solution becomes singular at the resonant surface.  

To describe the magnetic perturbations, it is convenient to introduce the flux function $\psi(x,z)$, defined such that the in-plane magnetic field is
\begin{equation}
\delta \mathbf{B}_\perp = \nabla \times (\psi \, \hat{y}) 
= \left(-\frac{\partial \psi}{\partial z}, \, 0, \, \frac{\partial \psi}{\partial x}\right).
\end{equation}
In this representation, the reconnecting magnetic field component is $\delta B_z = \partial_x \psi$.  

Solving the governing equation for $\psi$ in the outer layer provides the tearing stability parameter $\Delta'$, defined as the normalized jump in the logarithmic derivative of $\psi$ across the resonant surface,
\begin{equation}
\Delta' \equiv \left.\frac{1}{\psi}\frac{d\psi}{dz}\right|_{0^+}
- \left.\frac{1}{\psi}\frac{d\psi}{dz}\right|_{0^-}.
\end{equation}
Since $\delta B_z \propto \psi$ (for Fourier modes with $e^{i k_x x}$ dependence), this definition is equivalently expressed in terms of the reconnecting field perturbation,
\begin{equation}
  \label{eq:jump}
  \Delta' = \frac{1}{\delta B_z(0)}\left(\frac{d \delta B_z}{dz}\Big|_{0^+}
- \frac{d \delta B_z}{dz}\Big|_{0^-}\right).
\end{equation}
Both forms are identical, because $\delta B_z = i k_x \psi$ links the two variables.  

Physically, $\Delta'$ quantifies how sharply the magnetic perturbation bends across the singular layer. This parameter is then passed to the inner layer, a thin region around the current sheet where resistive effects are retained and magnetic reconnection occurs. Matching the inner solution to the outer one through $\Delta'$ yields a consistent global description of the instability and, in the constant-$\psi$ regime \citep{fkr1963}, leads directly to the estimation of the tearing mode growth rate.

The dispersion relation is finally obtained, taking into account that, for shorter wavelengths, the system enters the constant-$\psi$ regime, where the magnetic flux perturbation, $\psi$ (proportional to $\delta b_z$), remains nearly constant. Conversely, for longer wavelengths, it enters the nonconstant-$\psi$ regime, characterized by significant variation in $\psi$ and local dynamics. To find the maximum growth rate, $\sigma_\mathrm{max}$, we derive a scaling law by first calculating the growth rate $\sigma$ in each regime and then matching the two solutions to identify the transitional wavenumber between them. These steps are performed in the subsections below.

\subsection{Outer Region Solution}
\label{ssec:outer}

The derivation of the governing differential equation for the magnetic field perturbation $\delta B_z$ in the outer ideal region is obtained by ignoring resistivity and viscosity. Additionally, we consider $\Delta \beta_0 = 0$ in order to understand how the fluctuations drive pressure anisotropy from the isotropic equilibrium. The primary objective is to obtain a single, self-contained second-order differential equation for $\delta B_z$. The analysis begins with the $z$-component of the momentum equation, Eq.~(\ref{dvz}), and proceeds by expressing the velocity perturbation $\delta v_z$ in terms of $\delta B_z$ via the ideal version of Eq.~(\ref{dbz}). When this relation and its derivatives are substituted into the left-hand side of the momentum equation, the entire term is found to be of order $\mathcal{O}(\sigma^2)$:
\begin{equation}
\begin{split}
    - & \frac{\sigma^{2}}{k^2 B_0^2} \left[ \delta B_z'' - 2 \frac{B_0'}{B_0} \delta B_z' - \left(\frac{B_0''}{B_0} - 2\frac{B_0'^2}{B_0^2} + k^2 \right) \delta B_z \right] = \\
    & \delta B_z'' - \left( \frac{B_0''}{B_0} + k^2 \right) \delta B_z + i k \left( 2 B_0' \delta \Delta p + B_0 \delta \Delta p' \right).
\end{split}
\label{bz-relation}
\end{equation}

Because the tearing mode is a resistive instability, its growth rate $\sigma$ tends to zero as the resistivity $\eta \to 0$. Consequently, $\sigma$ is infinitesimally small, and the $\mathcal{O}(\sigma^2)$ terms on the left-hand side of Eq.~(\ref{bz-relation}) can be neglected, leading to an expression that couples the magnetic field perturbation $\delta B_z$ to the pressure perturbation $\delta \Delta p$ and its derivative $\delta \Delta p'$.

To eliminate the pressure terms, the remaining ideal equations are employed. Specifically, the $y$-component of the velocity perturbation, Eq.~(\ref{dvy}), is expressed in terms of $\delta \Delta p$ by combining the $y$-momentum and $y$-induction equations. In the small growth rate limit ($\sigma^2 \ll k^2 B_0^2$), this yields
\begin{equation}
\delta v_y \sim \frac{1}{i k}\,\frac{B_g}{B_0}\,\sigma\,\delta \Delta p.
\end{equation}

Substituting this result for $\delta v_y$ into the pressure evolution equation, Eq.~(\ref{dp_lin}), allows $\delta \Delta p$ to be expressed solely in terms of $\delta B_z$ and its derivatives. Differentiating this expression with respect to $z$ then provides the required relation for $\delta \Delta p'$.

The final step is to substitute the expressions for $\delta \Delta p$ and $\delta \Delta p'$ back into the momentum balance equation, Eq.~\ref{bz-relation}. After algebraic simplification, and making use of the equilibrium relations $B_g^2 = 1 - B_0^2$ and $B_g B'_g = -B_0 B'_0$, the equation reduces to the governing differential equation for the outer-region perturbation $\delta B_z$:
\begin{align}
  \label{eq:outer}
  \delta B''_z & + \frac{2 \tilde{\beta}}{1 + \tilde{\beta} B_g^{2}} B_0 B'_0 \, \delta B'_z \\ & - \left[ \frac{1 + \tilde{\beta} B_g^2}{1 + \tilde{\beta}} k^2 + \frac{B''_0}{B_0} + \frac{2 \tilde{\beta}}{1 + \tilde{\beta} B_g^{2}} B'^2_0 \right] \delta B_z = 0, \nonumber
\end{align}
where $\tilde{\beta} \equiv \frac{1}{2} \left( \gamma_\parallel + \gamma_\perp - 2 \right) \beta_0$.

When $\tilde{\beta} = 0$, i.e. when $\gamma_\parallel + \gamma_\perp = 2$, the additional pressure anisotropy contributions vanish, and the differential equation for the outer region, Eq.~(\ref{eq:outer}), reduces to the classical MHD form:
\begin{equation}
  \label{eq:mhd}
  \delta B''_z - \left[ k^2 + \frac{B''_0}{B_0} \right] \delta B_z = 0.
\end{equation}

After substituting the equilibrium fields given by the Harris current sheet profile, $B_0(z) = \tanh z$ and $B_g(z) = \sech z$, we obtain
\begin{align}
  \label{eq:outer-explicit}
  \delta B''_z & + \frac{2 \tilde{\beta} \tanh{z} \sech^2{z}}{1 + \tilde{\beta} \sech^2{z}} \delta B'_z \\
    & - \left[ \frac{1 + \tilde{\beta} \sech^2{z}}{1 + \tilde{\beta}} k^{2} - \frac{2\sech^{2} z}{1+\tilde{\beta} \sech^{2} z} \right] \delta B_z = 0 \nonumber
\end{align}

This equation highlights the effect of the modified plasma-$\beta$ parameter $\tilde{\beta}$ on the structure of the outer solution. The additional terms proportional to $\tilde{\beta}$ are localized around the current sheet center, since they are weighted by $\sech^2 z$, which decays exponentially for $|z|\gg 1$. Thus, $\tilde{\beta}$ modifies the dynamics primarily in the vicinity of the sheet, where magnetic field gradients are strongest. Far from the current sheet, $\sech^2 z \to 0$, so these corrections vanish, and the equation asymptotically reduces to the classical MHD form with constant coefficients.

More explicitly, in the asymptotic region ($|z|\to \infty$), Eq.~(\ref{eq:outer-explicit}) reduces to
\begin{equation}
  \delta B''_z - \frac{k^2}{1 + \tilde{\beta}} , \delta B_z = 0 ,
\end{equation}
so the perturbation behaves as $\delta B_z \sim e^{\pm \tfrac{k}{\sqrt{1+\tilde{\beta}}} z}$. This shows that the effective plasma-$\beta$ parameter $\tilde{\beta}$ rescales the decay length of the mode outside the current sheet: higher values of $\tilde{\beta}$ decrease the decay rate, broadening the eigenfunction. In contrast, near the sheet center ($z \approx 0$), the $\tilde{\beta}$–dependent terms in both the first-derivative coefficient and the effective potential dominate, altering the matching conditions that determine the tearing stability parameter $\Delta'$.

In summary, $\tilde{\beta}$ modifies the structure of the eigenfunction in two ways: (i) it introduces additional coupling through the first-derivative term, which enhances or suppresses the perturbation depending on the sign of $\tilde{\beta}$, and (ii) it changes the effective decay rate of $\delta B_z$ far from the sheet. When $\tilde{\beta} = 0$, all such corrections vanish, and Eq.~(\ref{eq:outer-explicit}) reduces to the standard outer-region tearing mode equation of classical resistive MHD.

The asymptotic analysis of Eq.~(\ref{eq:outer-explicit}) shows that, far from the current sheet, the perturbation decays exponentially with rate $\lambda = k/\sqrt{1+\tilde{\beta}}$. Thus, the general solution in this region is a linear combination, $\delta B_z(z) = c_1 e^{\lambda z} + c_2 e^{-\lambda z}$. Physical realizability requires the perturbation to vanish as $|z|\to \infty$, so the growing exponential must be excluded, leaving only the decaying branch. The key physical implication is that an increasing modified plasma-beta parameter $\tilde{\beta}$ reduces the decay rate $\lambda$, thereby broadening the spatial extent of the mode outside the current sheet.

Because the equilibrium fields satisfy $B_0(z)$ antisymmetric and $B_g(z)$ symmetric, the tearing perturbation $\delta B_z$ inherits a symmetric structure with respect to $z$. Consequently, $\delta B_z(z)$ is continuous across the midplane, but its derivative $\delta B'_z(z)$ exhibits opposite signs on either side, producing the well-known jump condition that defines the tearing stability parameter $\Delta'$. For this reason, it is sufficient to analyze the region $z>0$, with the solution for $z<0$ obtained by reflection symmetry. To incorporate the correct asymptotic decay directly into the solution, we introduce the ansatz
\begin{equation}
\label{eq:ansatz}
  \delta B_z(z) = e^{-\lambda z} f(z),
\end{equation}
where $f(z)$ approaches a constant as $z \to \infty$, and $\lambda = k/\sqrt{1+\tilde{\beta}}$ defines the inverse decay length.

By substituting the ansatz (Eq.~\ref{eq:ansatz}) together with its first- and second-order derivatives into Eq.~(\ref{eq:outer-explicit}), and using the identity $k^2 = \lambda^2 (1 + \tilde{\beta})$, the common exponential factor $e^{-\lambda z}$ cancels out. Collecting terms by the derivatives of $f(z)$ yields the following simplified differential equation:
\begin{align}
\label{eq:f_ode}
    & \left( 1 + \tilde{\beta} \sech^2{z} \right) f''(z) \\
    & + 2 \left[ \left( 1 + \tilde{\beta} \sech^{2} z \right) \left( \tanh{z} - \lambda \right) - \tanh{z} \right] f'(z) \nonumber \\
    & + \left\{ 2-\tilde{\beta} \lambda \left[ \lambda \left( 1+\tilde{\beta} \sech^{2} z\right) + 2\tanh z\right]\right\} \sech^2{z} \, f(z) = 0. \nonumber
\end{align}

In general, this second-order differential equation has no closed-form analytic solution. However, a complete exact solution is not required for the present analysis. To evaluate how the tearing stability parameter $\Delta'$ depends on the effective plasma-$\beta$ parameter $\tilde{\beta}$, it is sufficient to construct a local approximation near the sheet center ($z \simeq 0$). In the next subsection, we will therefore analyze the series solution of Eq.~\eqref{eq:f_ode} in order to derive the explicit expression for $\Delta'$.

\subsection{Derivation of the Tearing Stability Index $\Delta'$}
\label{ssec:jump}

To obtain the tearing stability parameter $\Delta'$ in the gyrotropic MHD framework, we first transform Eq.~(\ref{eq:f_ode}) into a more convenient form by introducing the substitution $\tanh z = u$, which implies $\sech^2 z = 1 - u^2$. This change of variable eliminates the explicit hyperbolic functions and converts the equation into polynomial coefficients in $u$.  

The derivatives transform as $f'(z) = (1-u^2) f'(u)$ and $f''(z) = (1-u^2)^2 f''(u) - 2u(1-u^2)f'(u)$. Substituting these into Eq.~(\ref{eq:f_ode}) gives the equivalent differential equation for $f(u)$:
\begin{align}
\label{eq:f_ode_u}
& \left[ 1+\tilde{\beta} \left( 1-u^{2}\right)\right]\left( 1-u^{2}\right) f''( u) \\
& +\left\{2\tilde{\beta} u\left( 1-u^{2}\right) -2( \lambda +u)\left[ 1+\tilde{\beta} \left( 1-u^{2}\right)\right]\right\} f'( u) \nonumber \\
& +\left\{2-\tilde{\beta} \lambda \left[ \lambda \left[ 1+\tilde{\beta} \left( 1-u^{2}\right)\right] +2u\right]\right\} f( u) = 0. \nonumber
\end{align}

Because the tearing solution $\delta B_z(z)$ is symmetric, its derivative is antisymmetric, which implies $\delta B'_z(0^-) = -\delta B'_z(0^+)$. This property simplifies the jump condition, $\Delta'$, to
\begin{equation}
  \Delta' = 2 \frac{\delta B_z' ( 0^+) }{\delta B_z(0)}.
\end{equation}
Using the ansatz (Eq.~\ref{eq:ansatz}), with $\delta B_z'(z) = e^{-\lambda z}[f'(z)-\lambda f(z)]$, we evaluate this expression at $z=0$ and obtain
\begin{equation}
  \Delta' = 2 \left[ \frac{f'(0^+)}{f(0)} - \lambda \right].
\end{equation}
Thus, the tearing parameter is determined by the slope-to-value ratio of $f(z)$ at the sheet center, corrected by the decay rate $\lambda$.

To compute this ratio, we expand $f(u)$ in a power series around $u=0$:
\begin{equation}
    f(u) = \sum_{m=0}^{\infty} a_m u^m .
    \label{series}
\end{equation}
At the origin, $f(0)=a_0$ and $f'(0)=a_1$, so
\begin{equation}
    \Delta' = 2\left( \frac{a_1}{a_0} - \lambda \right).
    \label{eq:jump_approx}
\end{equation}

The coefficients of the series are constrained by Eq.~(\ref{eq:f_ode_u}). Evaluating the equation at $u=0$ gives
\begin{equation}
    2(1+\tilde{\beta})a_{2}-2\lambda(1+\tilde{\beta})a_{1}+[2-\tilde{\beta}\lambda^{2}(1+\tilde{\beta})]a_{0}=0.
\label{eq:coeffs_relation}
\end{equation}
From this relation,
\begin{equation}
    \frac{a_{1}}{a_{0}}=\left(\frac{1}{1+\tilde{\beta}}+\frac{a_{2}}{a_{0}}\right)\frac{1}{\lambda}-\frac{1}{2}\tilde{\beta}\lambda.
\end{equation}
Substituting into Eq.~(\ref{eq:jump_approx}) yields the general form of the tearing parameter,
\begin{equation}
    \Delta' = \left( \frac{1}{1+\tilde{\beta}} + \frac{a_2}{a_0} \right) \frac{2}{\lambda} - (2 + \tilde{\beta})\lambda ,
\label{eq:Delta_full}
\end{equation}
which shows that $\Delta'$ depends explicitly on $\tilde{\beta}$, the wavenumber $k$ (through $\lambda$), and the ratio $a_2/a_0$ that encodes the curvature of the solution at the sheet center.

The series expansion in Eq.~(\ref{series}) generates a recursive relation for all higher-order coefficients $a_{\rm m}$ with $m \geq 2$. In practice, the series can be truncated by setting all higher-order coefficients to zero, which yields a closed system for the ratios $a_1/a_0$ and $a_2/a_0$.

Our numerical investigation of the differential equation (\ref{eq:f_ode_u}) shows, however, that $a_2$ is much smaller in magnitude compared to $a_1$, and its influence can therefore be neglected. Nevertheless, we adopt the approximate expression
\begin{equation}
\frac{a_2}{a_0} \approx - \frac{1}{2} \tilde{\beta} \lambda^2,
\end{equation}
which perfectly recovers the stability criterion discussed below.

Substituting this approximation for $a_2/a_0$ into the general expression for $\Delta'$ gives the following practical form of the jump condition:
\begin{equation}
  \Delta' \approx 2 \left[ \frac{1}{(1 + \tilde{\beta}) \lambda} - (1 + \tilde{\beta}) \lambda \right].
\label{eq:jump_final}
\end{equation}

The stability of the tearing mode can now be assessed directly from the approximate expression for $\Delta'$, Eq.~(\ref{eq:jump_final}). The mode is stable when $\Delta' \leq 0$ and unstable when $\Delta' > 0$. From the explicit form of $\Delta'$, instability occurs when the decay rate $\lambda$ satisfies
\begin{equation}
    \lambda < \frac{1}{1 + \tilde{\beta}},
\end{equation}
or, equivalently, when the wavenumber $k$ lies below the threshold
\begin{equation}
    k < \frac{1}{\sqrt{1 + \tilde{\beta}}},
\end{equation}

Thus, the presence of the effective plasma-$\beta$ parameter $\tilde{\beta}$ shifts the marginal stability boundary by reducing the range of unstable wavenumbers. In the classical MHD limit $\beta_0 \to 0$ (or when $\gamma_\parallel + \gamma_\perp = 2$), this condition reduces to the well-known result $k < 1$ for the onset of tearing instability.

\subsection{Inner Region Dynamics and Term Balances}
\label{ssec:inner}

We now turn to the inner region in order to estimate the scaling of the tearing mode growth rate. Our goal here is not to solve the full set of equations but rather to identify the dominant balances that control the instability and to obtain scaling relations for the maximum growth rate $\sigma_\mathrm{max}$ and the corresponding most unstable wavenumber $k_\mathrm{max}$.  

In the thin inner layer ($|z| \ll 1$), the resistive sublayer has a width $h \ll a$ (recall that the current sheet thickness is normalized to $a=1$). Within this region, second derivatives across the sheet dominate over terms involving the long-wavelength parallel variation:
\begin{equation}
  \delta B_z'' \gg k^2 \delta B_z, \qquad 
  \delta v_z'' \gg k^2 \delta v_z .
\end{equation}
Moreover, near the origin, the equilibrium field is approximately linear, $B_0(z)=\tanh z \simeq z$, so the curvature term $B_0'' \delta B_z$ is negligible compared to $B_0 \delta B_z''$.  

The momentum equation also contains terms proportional to the perturbed pressure difference, $\delta \Delta p$ and $\delta \Delta p'$, weighted by $k^2 B_0$. In the inner region, these terms are subdominant: since $k \ll 1$ and $B_0(z)\sim z \ll 1$, they are small compared to the magnetic tension term $B_0 \delta B_z''$ and can therefore be neglected. Physically, this reflects the fact that the very narrow resistive layer is dominated by magnetic tension and diffusion, with pressure anisotropy effects entering only through the outer boundary condition $\Delta'$.  

With these simplifications, the inner-layer equations reduce to
\begin{align}
  \label{class_tearing}
  \sigma \, \delta v_z'' &\simeq i k\, B_0 \,\delta B_z'', \\
  \sigma \, \delta B_z &\simeq i k B_0 \,\delta v_z + \eta \,\delta B_z'' .
\end{align}

These governing equations are identical to those of the classical resistive MHD tearing instability. The gyrotropic corrections do not modify the inner-layer dynamics directly; instead, their influence is fully contained in the stability parameter $\Delta'$ determined by the outer solution.  

Following the standard inner-layer analyses \cite[see, e.g.,][]{fkr1963, 1976SvJPP...2..533C, 1993noma.book.....B, 2007PhPl...14j0703L, Somov2013}, one can identify the essential balances between terms. In particular, the inertial term balances magnetic tension, 
\begin{equation}
  \sigma \, \delta v_z'' \sim k B_0 \, \delta B_z'', 
\end{equation}
while inductive growth balances convective drive and resistive diffusion,  
\begin{equation}
  \sigma \, \delta B_z \sim k B_0 \, \delta v_z \sim \eta \,\delta B_z'' .
\end{equation}

These relations summarize the three-way balance of induction, convection, and diffusion that governs the resistive sublayer. They form the basis for estimating the tearing mode growth rate in the two classical regimes: constant-$\psi$ and nonconstant-$\psi$. Importantly, only the constant-$\psi$ regime depends on $\Delta'$, and hence on the effective plasma-$\beta$ parameter $\tilde{\beta}$ through the outer solution.

\begin{figure*}[t]
\centering
\includegraphics[width=\linewidth]{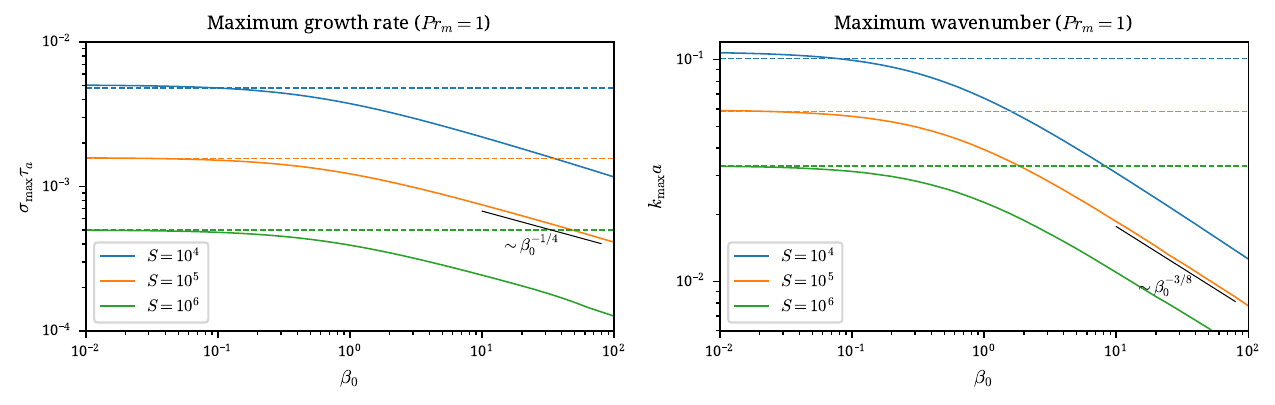}
\caption{Maximum growth rate (left) and the corresponding maximum wavenumber (right) of the tearing mode instability as functions of the equilibrium plasma-$\beta$ parameter for $\Delta \beta=0$ for different Lundquist numbers $S = 10^4$, $10^5$, and $10^6$, with $Pr_{\rm m}=1$. For the plasma-$\beta$ dependence, the analysis considers both the MHD framework (dashed) and the nonideal gyrotropic MHD model (solid).}
\label{fig:beta-dependence}
\end{figure*}

\begin{figure*}
\centering
\includegraphics[width=\linewidth]{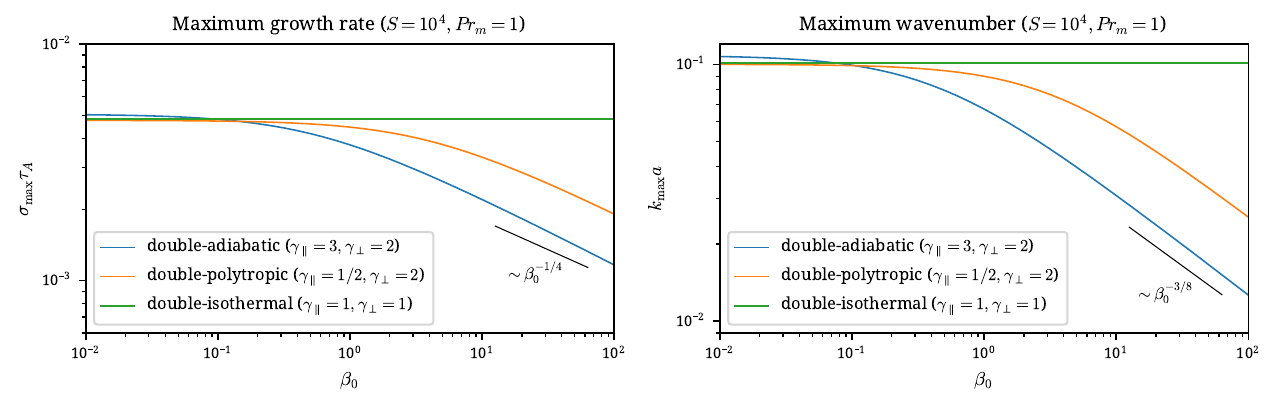}
\caption{Comparison of maximum growth rates and wavenumbers for different polytropic closures in the nonideal gyrotropic MHD model. Left panel: the normalized maximum growth rate ($\sigma_\mathrm{max}\tau_A$) as a function of plasma-$\beta$. Right panel: the wavenumber corresponding to the maximum growth rate ($k_\mathrm{max}a$) as a function of plasma-$\beta$. The results are shown for a Lundquist number $S=10^4$ and a magnetic Prandtl number $Pr_{\rm m}=1$. The lines represent three distinct polytropic closures: double-adiabatic ($\gamma_\parallel=3, \gamma_\perp=2$; blue), double-polytropic ($\gamma_\parallel=1/2, \gamma_\perp=2$; orange), and double-isothermal ($\gamma_\parallel=1, \gamma_\perp=1$; green). The figure demonstrates that the growth rate's dependence on $\beta$ emerges in nonisothermal models, with a notable suppression in high-$\beta$ regimes, whereas the isothermal model remains independent of $\beta$, consistent with classical MHD theory. The black line indicates the reference slope $\sim\beta^{-1/4}$ for the growth rate scaling at high $\beta$.}
\label{fig:different_gammas}
\end{figure*}

\subsection{Growth Rate and $\tilde{\beta}$ Dependence}
\label{ssec:growth_rate}

From the outer-region analysis, the stability parameter $\Delta'$ (Eq.~\ref{eq:jump_final}) takes the general form
\begin{equation}
  \Delta' \sim \frac{1}{k \sqrt{1+\tilde{\beta}}} - k \sqrt{1+\tilde{\beta}}.
\end{equation}
For long wavelengths ($k \ll 1/\sqrt{1+\tilde{\beta}}$), the first term dominates and $\Delta'$ is large and positive, while for shorter wavelengths, the second term becomes important and eventually drives $\Delta'$ negative, recovering the classical stability threshold $k \simeq 1/\sqrt{1+\tilde{\beta}}$.  

\noindent
\textit{Constant-$\psi$ regime.}
In the constant-$\psi$ limit \citep{fkr1963}, the growth rate satisfies
\begin{equation}
    \sigma^5 \sim (k B_0')^2 \Delta'^4 \eta^3 .
\end{equation}
Substituting $\Delta'$ and using $\eta \sim S^{-1}$ gives
\begin{equation}
    \sigma^5 \sim (k B_0')^2 \left[ \frac{1}{k \sqrt{1+\tilde{\beta}}} - k \sqrt{1+\tilde{\beta}} \right]^4 S^{-3}.
\end{equation}
Thus, the $\tilde{\beta}$ dependence enters only through the combination $k \sqrt{1+\tilde{\beta}}$ that controls both the magnitude and the sign of $\Delta'$.  

\noindent
\textit{Nonconstant-$\psi$ regime.}  
In the nonconstant-$\psi$ limit \citep{1976SvJPP...2..533C}, the growth rate scales as
\begin{equation}
    \sigma^3 \sim k^2 S^{-1},
\end{equation}
which is independent of $\Delta'$ and hence of $\tilde{\beta}$.  

\noindent
\textit{Maximum growth rate.}  
Since the constant-$\psi$ growth rate decreases with $k$ while the nonconstant-$\psi$ growth rate increases with $k$, the two curves intersect at a critical wavenumber $k_m$ that defines the most unstable mode. Equating the two asymptotic scalings gives
\begin{equation}
    k_m^{2/3} S^{-1/3} \sim (k_m B_0')^{2/5} \Delta'^{4/5} S^{-3/5}.
\end{equation}
Using the expression for $\Delta'$ valid at the transition point ($k_m \lesssim 1/\sqrt{1+\tilde{\beta}}$), one finds
\begin{equation}
    k_m \sim (1+\tilde{\beta})^{-3/8} S^{-1/4}.
\end{equation}
Substituting this into the nonconstant-$\psi$ scaling law yields the maximum growth rate,
\begin{equation}
    \sigma_m \sim (1+\tilde{\beta})^{-1/4} S^{-1/2}.
\end{equation}

Expressing the result in terms of the adiabatic indices gives
\begin{equation}
    \sigma_m \sim \left[ 1 + \tfrac{1}{2}(\gamma_\parallel + \gamma_\perp - 2)\beta_0 \right]^{-1/4} S^{-1/2}.
\end{equation}
For the double-adiabatic case ($\gamma_\parallel=3$, $\gamma_\perp=2$),
\begin{equation}
  \sigma_m \sim \left( 1 + \tfrac{3}{2}\beta_0 \right)^{-1/4} S^{-1/2}.
\end{equation}

The tearing instability in the gyrotropic MHD framework thus preserves the classical $S^{-1/2}$ scaling for the fastest-growing mode, but the growth rate is reduced by a factor $(1+\tilde{\beta})^{-1/4}$ relative to isotropic MHD. The effect of $\tilde{\beta}$ is therefore stabilizing: increasing effective plasma-$\beta$ both narrows the range of unstable wavenumbers and decreases the maximum growth rate.

\begin{figure*}
\centering
\includegraphics[width=\linewidth]{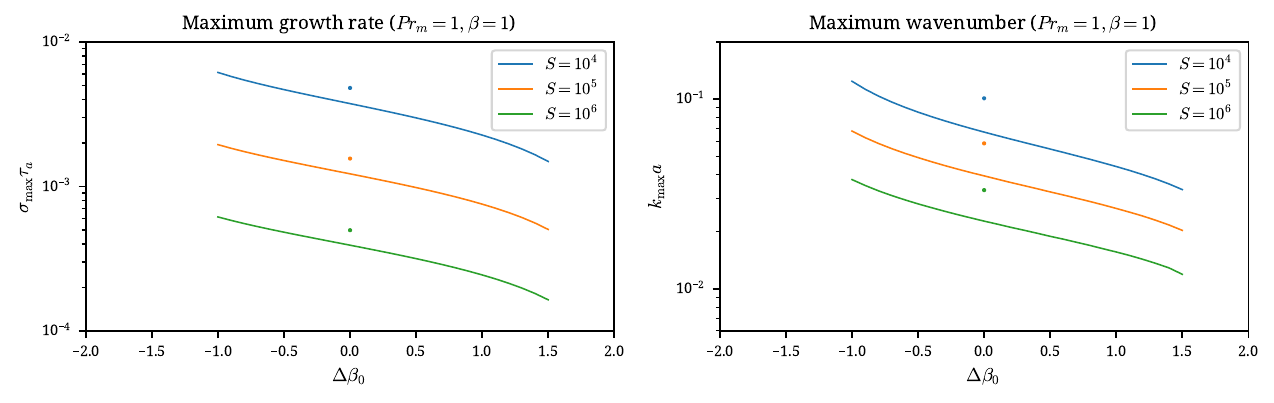}
\includegraphics[width=\linewidth]{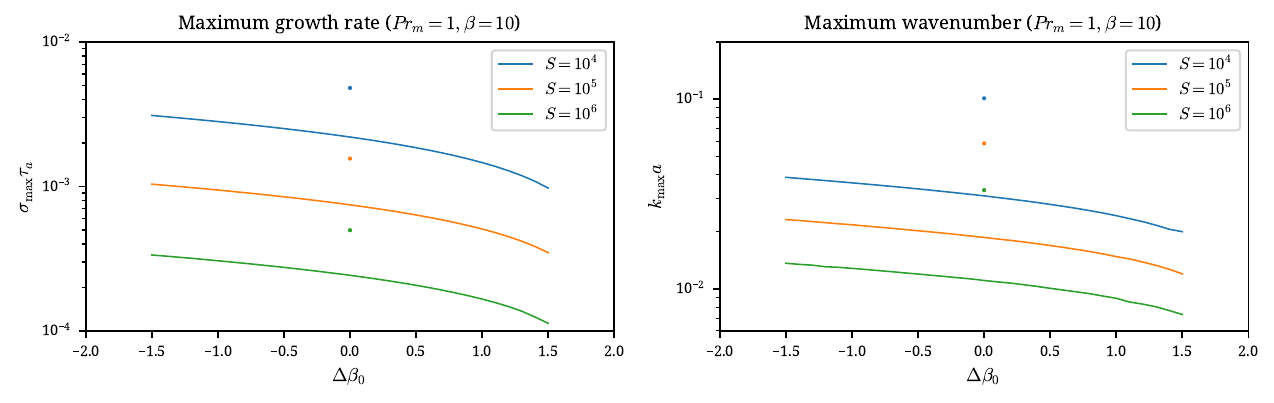}
\caption{Maximum growth rate (left) and the corresponding maximum wavenumber (right) of the tearing mode instability as functions of the equilibrium pressure anisotropy parameter $\Delta\beta_0$, for $\beta = 1$ and $10$ (upper and lower rows, respectively) for different Lundquist numbers $S = 10^4$, $10^5$, and $10^6$, with $Pr_{\rm m}=1$. The points at $\Delta \beta_0 = 0$ represent $\sigma_\mathrm{max} \tau_a$ and $k_\mathrm{max} a$ obtained from the MHD approximation.}
\label{fig:anisotropy-dependence}
\end{figure*}

\section{Numerical Solutions}
\label{sec:numerical}

To validate the analytical scaling laws derived in Section~\ref{sec:theory} we analyze numerically the stability of the linearized nonideal gyrotropic MHD equations. We employ the Pseudo-Spectral Eigenvalue Calculator with an Automated Solver (\texttt{PSECAS}) framework \citep{berlok2019}, which efficiently solves eigenvalue problems arising in fluid and plasma stability analyses. First, we express the linearized equations (Eqs. \ref{dvy}-\ref{dp_lin}) in a matrix form, where perturbations are expanded as normal modes in the form described above, reducing the problem to a set of coupled ordinary differential equations (ODEs) in $z$. \texttt{PSECAS} discretizes these ODEs using a pseudospectral collocation method on a nonuniform grid optimized for resolving steep gradients, mapping the problem onto a generalized eigenvalue problem. It then systematically increases the resolution of the eigenproblem matrices and verifies the convergence of the computed growth rates to ensure numerical accuracy. The resulting eigenvalue spectrum provides the growth rates ($\sigma$) and mode structures of tearing instability under different equilibrium conditions, allowing us to systematically investigate the effects of plasma-$\beta$, pressure anisotropy, viscosity, and resistivity on stability in the nonideal gyrotropic MHD framework.

Our numerical study systematically investigates the influence of the key equilibrium parameters: the plasma-$\beta$ ($\beta_0$), the equilibrium pressure anisotropy ($\Delta\beta_0$), the Lundquist number ($S$), and the magnetic Prandtl number ($Pr_{\rm m}$). The results robustly confirm our central analytical prediction: the emergence of a stabilizing $\beta$-dependence in the tearing mode growth rate, a feature absent in classical MHD.

Figure \ref{fig:beta-dependence} (left panel) illustrates this key finding. For an initially isotropic equilibrium ($\Delta\beta_0=0$), the numerically computed maximum growth rate ($\sigma_\mathrm{max}$) closely follows the analytical scaling $\sigma_\mathrm{max}\tau_a \propto \beta_0^{-1/4}$ in the high-$\beta$ regime ($\beta_0 \gg 1$; solid lines). This contrasts sharply with the classical MHD result, which shows no dependence on $\beta_0$ (dashed lines). Furthermore, the wavenumber of the most unstable mode ($k_\mathrm{max}$) shifts to smaller values as $\beta_0$ increases, closely following the analytical prediction, $k_\mathrm{max} a \propto \beta_0^{-3/8}$. This result indicates that thermal pressure restricts the instability to larger spatial scales. As shown in Figure \ref{fig:different_gammas}, this behavior is consistently observed for different nonisothermal polytropic closures (e.g., double-adiabatic), while the double-isothermal case ($\gamma_\parallel=\gamma_{\perp}=1$) correctly recovers the classical MHD limit where the growth rate is independent of $\beta_0$.

The numerical solutions also confirm the influence of equilibrium anisotropy, $\Delta\beta_0$. As shown in Figure \ref{fig:anisotropy-dependence} (left panels), a positive initial anisotropy ($\Delta\beta_0 > 0$, or $p_{\parallel,0} > p_{\perp,0}$) enhances the suppression of the tearing mode, with the wavenumber of the fastest-growing modes shifting toward larger scales. Conversely, a negative anisotropy ($\Delta\beta_0 < 0$, or $p_{\parallel,0} < p_{\perp,0}$) can counteract the stabilizing effect of plasma-$\beta$, potentially leading to growth rates that exceed MHD predictions in low-$\beta$ regimes, with the corresponding wavenumbers shifting toward shorter scales (compare the top and bottom panels for $\beta_0 = 1$ and $10$, respectively). These numerical results provide strong evidence for the validity of our analytical model and underscore the critical role of pressure anisotropy dynamics in governing tearing instability in weakly collisional plasmas.

\section{Discussion}

\label{sec:discussion}

The analytical and numerical results presented in this work reveal a novel scaling of the tearing mode instability with plasma-$\beta$ in weakly collisional environments, challenging the long-standing paradigm based on resistive MHD. The primary finding -- that the maximum growth rate scales as $\sigma_\mathrm{max} \tau_a\propto \beta^{-1/4}$ for high $\beta$ -- stems directly from relaxing the assumption of strict pressure isotropy. Although our main numerical results are presented for the double-adiabatic indices ($\gamma_\parallel = 3$, $\gamma_\perp = 2$), for direct comparison with classical limits, we have derived the maximum growth rate dependence on plasma-$\beta$ for the general case of arbitrary adiabatic indices $\gamma_\parallel$ and $\gamma_\perp$. Our analytical and numerical analysis confirms that, for any values of these indices -- provided they are not double-isothermal ($\gamma_\parallel = \gamma_\perp = 1$) -- the growth rate depends on plasma-$\beta$. This $\beta$-dependence of the tearing growth rate therefore persists across a wide range of closures, although the specific scaling coefficients may vary.

The stabilizing effect of plasma-$\beta$ on the tearing instability, unveiled in this work, adds a new dimension to our understanding of reconnection in anisotropic plasmas. Early theoretical considerations by \cite{1973NASSP.342..357S, Sonnerup1974} already suggested that tearing in symmetric current sheets should be suppressed at high $\beta$ due to a gyroresistive mechanism, predicting a critical threshold $\beta_p \lesssim 1$ for reconnection onset. His argument did not invoke gyrotropic pressure equations but rather followed from a consistency requirement of resistive MHD: the diffusion region thickness must exceed the ion gyroradius, which in turn scales with $\sqrt{\beta}$. At high $\beta$, this condition can only be met if the effective resistivity is strongly enhanced above the Spitzer value through anomalous or gyroresistive processes, which he noted are operative only when the inflow $\beta$ remains small. Subsequent studies shifted the focus toward the destabilizing role of pressure anisotropy, with the kinetic Vlasov treatment by \cite{Chen1984PhFl...27.1198C} demonstrating that ion anisotropy ($T_{i\perp} > T_{i\parallel}$) could enhance tearing growth rates by up to an order of magnitude, while also revealing the limitations of the conventional two-region matching analysis by identifying a crucial intermediate region governed by axis-crossing ion orbits. From a fluid perspective, the double-polytropic MHD analyses of \cite{2002GeoRL..29.1815C, 2003PhPl...10.3813C} extended this line of inquiry, showing that the tearing growth rate increases with both the anisotropy ratio $p_{\perp}/p_{\parallel}$ and plasma-$\beta$, which they attributed to the system more easily satisfying the mirror instability criterion at high $\beta$.

Our results present a contrasting and unifying refinement to this picture. While we recover the stabilizing influence of $p_{\parallel,0} > p_{\perp,0}$ consistent with earlier findings, our central result is that in a weakly collisional, initially isotropic plasma, a stabilizing pressure anisotropy is generated dynamically, and its effect strengthens with increasing $\beta$, leading to the suppression $\sigma_{\max}\tau_A \propto \beta^{-1/4}$ for $\beta \gg 1$. This scaling not only confirms and extends the early prediction of Sonnerup's gyroresistive framework but also clarifies the apparent contradiction with the conclusions of \cite{2002GeoRL..29.1815C, 2003PhPl...10.3813C}. The discrepancy arises from their adoption of a double-polytropic closure that included resistive diffusion in the induction equation but neglected the corresponding ohmic heating in the parallel and perpendicular pressures, thereby violating energy conservation. Without this heating channel, their model underestimated the stabilizing $\beta$-dependent feedback inherent to anisotropy fluctuations and instead emphasized mirror-mode coupling as the dominant effect, leading them to predict reconnection enhancement at large $\beta$. By contrast, our consistent nonideal gyrotropic formulation restores energy balance and isolates the fundamental thermodynamic damping mechanism intrinsic to high-$\beta$ plasmas, establishing the $\beta^{-1/4}$ law as the correct asymptotic behavior of the tearing instability.

\subsection{Physical Interpretation of the $\beta$-dependence}

In the gyrotropic MHD framework, even an initially isotropic plasma ($\Delta\beta_0=0$) is free to develop pressure anisotropy dynamically. Perturbations in the velocity and magnetic fields ($\delta\mathbf{v}$, $\delta\mathbf{B}$) induce fluctuations in the pressure anisotropy ($\delta\Delta p$). As shown in our derivation (Eq.~\ref{dp_lin}), the evolution of $\delta\Delta p$ is explicitly coupled to the plasma-$\beta$. This fluctuating pressure anisotropy introduces a restoring force in the momentum equation that acts to suppress the instability, an effect that becomes more pronounced at higher $\beta$. This physical mechanism, which is absent by definition in standard MHD, is responsible for the stabilizing $\beta$-dependence found in this work. The shift of the instability to longer wavelengths at high $\beta$ can be understood as a consequence of increased thermal pressure, which effectively "stiffens" the plasma and limits the tearing mode to larger, less constrained spatial scales.

\subsection{Reconnection in Weakly Collisional Plasmas}

Our findings have significant implications for magnetic reconnection in a variety of astrophysical settings characterized by high-$\beta$ and weakly collisional plasmas, such as the solar corona, heliospheric current sheets, planetary magnetospheres, and the intracluster medium. The conventional view of fast reconnection is often mediated by the formation of plasmoids, whose onset is governed by the tearing instability. The $\beta$-dependent suppression of the linear tearing mode suggests that the conditions for the onset of plasmoid-mediated reconnection may be more restrictive in high-$\beta$ environments than previously thought based on resistive MHD models.

Furthermore, the sensitivity of the growth rate to the sign of the equilibrium anisotropy ($\Delta\beta_0$) provides a more nuanced picture. Environments where kinetic processes naturally drive the plasma toward $p_\perp > p_\parallel$ (e.g., regions of magnetic field compression) could experience enhanced tearing growth rates compared to MHD predictions, potentially accelerating the onset of reconnection. Conversely, in regions where $p_\parallel > p_\perp$ (e.g., due to magnetic field line stretching), the instability would be significantly suppressed. These results underscore the necessity of incorporating pressure anisotropy effects into large-scale reconnection models to accurately represent astrophysical phenomena.

Our results on the $\beta$-dependence of the tearing instability can be placed in the broader context of recent studies of reconnection in weakly collisional plasmas. \cite{2021ApJ...912..152L} showed, for antiparallel collisionless reconnection, that when the upstream ion plasma-$\beta$ exceeds unity, ion demagnetization in the exhaust enhances perpendicular heating and produces a normal stress gradient that limits the outflow speed, thereby reducing the reconnection rate. \cite{2024ApJ...977..218G} extended this framework to strong guide field reconnection, demonstrating that the total ion plasma-$\beta$ controls the transition from Alfvénic to sub-Alfvénic outflows through a similar back-pressure mechanism. While these works focus on the nonlinear steady-state phase of reconnection rather than the linear development of tearing modes, the physical picture they present suggests a possible link: in high-$\beta$ regimes, the same outflow-limiting processes identified in their studies could act during the nonlinear evolution of tearing-unstable current sheets, thereby influencing the effective reconnection rate and energy conversion efficiency in weakly collisional plasmas.

\subsection{Limitations and Open Questions}

While this study provides new fundamental insights, it is important to acknowledge its limitations. Our analysis is based on a linear, single-fluid model and does not capture the full, scale-dependent nature of magnetic reconnection. At kinetic scales (i.e., below the ion skin depth), two-fluid effects, Hall physics, and electron pressure gradients become dominant and are essential for enabling fast reconnection rates.

The transition from the large, macroscopic scales governed by anisotropic fluid dynamics to the small, kinetic scales where reconnection ultimately occurs remains a critical open question. How does the $\beta$-dependent suppression of the tearing mode at large scales influence the triggering of kinetic-scale reconnection? Moreover, in many astrophysical systems, turbulence is believed to play a crucial role in enabling fast reconnection. Future work should therefore aim to bridge these regimes, investigating the interplay between pressure anisotropy, turbulence, and kinetic physics in a unified framework.

\section{Conclusions}

\label{sec:conclusions}

In this work, we investigated the linear stability of the tearing mode in weakly collisional, gyrotropic plasmas using the gyrotropic MHD framework. Through a combination of analytical theory and numerical simulations, we extended classical MHD analyses to include the effects of pressure anisotropy. Our main conclusions are as follows.

\begin{enumerate}
    \item We unveiled a previously unknown dependence of the tearing instability growth rate on the plasma-$\beta$. In the gyrotropic MHD framework, the maximum growth rate is suppressed in high-$\beta$ plasmas, following an asymptotic scaling law of $\sigma_\mathrm{max}\tau_a \propto \beta^{-1/4}$. This effect arises from dynamically generated pressure anisotropy fluctuations that introduce a stabilizing restoring force, a mechanism entirely absent in standard resistive MHD.

    \item Our analytical derivation and numerical results both confirm that the scaling of the maximum growth rate with the Lundquist number ($S$) and plasma-$\beta$ is given by $\sigma_m \propto S^{-1/2} ( 1 + \tilde{\beta} )^{-1/4}$, while the corresponding wavenumber scales as $k_\mathrm{max} a \propto S^{-1/4} ( 1 + \tilde{\beta} )^{-3/8}$, where $\tilde{\beta} = \frac{1}{2} \left( \gamma_\parallel + \gamma_\perp - 2 \right) \beta_0$. These formulae generalize the classical \cite{fkr1963} scaling by incorporating the stabilizing effect of thermal pressure in a weakly collisional plasma.

    \item The stability of the tearing mode is highly sensitive to the equilibrium pressure anisotropy. When the parallel pressure dominates ($p_{\parallel,0} > p_{\perp,0}$), the instability is further suppressed. Conversely, when the perpendicular pressure dominates ($p_{\parallel,0} < p_{\perp,0}$), the stabilizing effect of plasma-$\beta$ can be counteracted, and for low-$\beta$ conditions, the growth rate can even exceed that predicted by classical MHD.

    \item In high-$\beta$ regimes, the wavenumber of the fastest-growing mode decreases, shifting the tearing instability to larger spatial scales. This suggests that in environments such as the intracluster medium or parts of the solar corona, magnetic reconnection via the tearing mode would preferentially occur at macroscopic wavelengths.
\end{enumerate}

These findings challenge the universal applicability of reconnection models that neglect pressure anisotropy. They provide a new theoretical foundation for understanding how magnetic energy is released in high-$\beta$, weakly collisional plasmas across the Universe. Future research should focus on integrating these anisotropic effects into nonlinear and multiscale models to build a more complete picture of magnetic reconnection.

\begin{acknowledgments}
The authors acknowledge support from FAPESP (grants 2013/10559-5, 2021/02120-0, 2021/06502-4 2022/03972-2,, and 2024/16327-3) and CAPES (88887.951845/2024-00). The simulations presented in this work were performed using the clusters of the Group of Theoretical Astrophysics at EACH-USP (Hydra HPC), which was acquired with support from FAPESP (grants 2013/04073-2 and 2022/03972-2).
\end{acknowledgments}

\newpage

\appendix

\section{Linearization of the Nonideal Gyrotropic MHD Equations}
\label{app:linearization}

In this Appendix, we present the detailed linearization of the nonideal gyrotropic MHD equations, which govern the evolution of an anisotropic plasma in the presence of a guiding magnetic field. The derivation follows the standard approach of expanding the equations around an equilibrium configuration and retaining only first-order perturbations. Compared to the standard MHD framework, the key difference in our nonideal gyrotropic-MHD model lies in the inclusion of the gyrotropic pressure tensor, introducing additional terms that influence the system's stability. The Hall term, which represents electron--ion decoupling, becomes critical only when resolving the inner diffusion layer at kinetic scales \cite{2019ApJ...885...56P}. Since our analysis is confined to the linear growth phase and focuses on the global scaling of the instability, such fine-scale effects are beyond the scope of this study. The nonideal gyrotropic MHD equations employed here provide a consistent framework for weakly collisional plasmas, especially in regimes where pressure anisotropy develops naturally and influences the macroscopic dynamics.

We begin by linearizing the momentum equation, highlighting the effects of pressure anisotropy. Subsequently, we derive the linearized pressure difference equation, accounting for contributions from equilibrium anisotropy, ohmic heating, and viscosity. Finally, we obtain the full set of coupled linearized equations governing the evolution of velocity, magnetic field, and pressure perturbations, which serve as the basis for the stability analysis presented in the main text.
 
When considering the momentum equation, the difference between the MHD and gyrotropic MHD approximations stems from the introduction of the pressure tensor in the latter. This causes the pressure term to become
\begin{equation}
\nabla \cdot \left[ p_\perp \mathbf{I} + \left( p_\parallel - p_\perp \right) \mathbf{b} \mathbf{b} \right],
\end{equation}
where $p_\parallel$ and $p_\perp$ are the parallel and perpendicular components of the pressure tensor and $\mathbf{b} \equiv \mathbf{B} / \left| \mathbf{B} \right|$. Applying vector identities, the anisotropic component of the above term can be expressed as
\begin{equation}
\left( \mathbf{B} \cdot \nabla \right) \left( \frac{\Delta p}{\left| \mathbf{B} \right|^2} \mathbf{B} \right),
\end{equation}
where $\Delta p \equiv p_\parallel - p_\perp$.

Since our equilibrium field has unit magnitude everywhere, $\left| \mathbf{B}_0 \right| = 1$, and applying the Taylor expansion of the inverse magnetic field magnitude for small magnetic field perturbation, $\left| \delta \mathbf{B} \right| \ll 1$,
\begin{equation}
\frac{1}{\left| \mathbf{B} \right|^2} = \frac{1}{\left( \mathbf{B}_0 + \delta \mathbf{B} \right) \cdot \left( \mathbf{B}_0 + \delta \mathbf{B} \right)} \approx 1 - 2 \left( \mathbf{B}_0 \cdot \delta \mathbf{B} \right),
\end{equation}
the resulting linearized pressure anisotropy term is
\begin{equation}
\left( \mathbf{B} \cdot \nabla \right) \left( \frac{\Delta p}{\left| \mathbf{B} \right|^2} \mathbf{B} \right) \approx \left( \mathbf{B}_0 \cdot \nabla \right) \left( \mathbf{B}_0 \delta \Delta p \right) + \Delta p_0 \left\{ \left( \mathbf{B}_0 \cdot \nabla \right) \delta \mathbf{B} + \left( \delta \mathbf{B} \cdot \nabla \right) \mathbf{B}_0 \right. - 2 \left. \left( \mathbf{B}_0 \cdot \nabla \right) \left[ \mathbf{B}_0 \left( \mathbf{B}_0 \cdot \delta \mathbf{B} \right) \right] \right\}
\end{equation}

In order to eliminate the pressure contribution in the linearized MHD equations, the momentum equation is transformed into the vorticity equation by applying the curl operator to all terms. In the gyrotropic MHD equations, the curl removes the contribution from the perpendicular pressure; however, the anisotropic pressure component does not vanish. Therefore, the curl operator must be applied to the linearized anisotropic pressure term, as shown above.

We observe two principal contributions to this resulting term: the first term on the right-hand side, which describes the effects of pressure anisotropy perturbation, and the remaining terms, which are nonnegligible when the equilibrium exhibits pressure anisotropy. Consequently, it is necessary to consider the pressure equations. However, because only pressure differences are relevant, we linearize the equation for the pressure difference, as shown below, rather than linearizing equations for parallel and perpendicular pressure separately:
\begin{align}
\frac{\partial \Delta p}{\partial t} = & - \left( \mathbf{v} \cdot \nabla \right) \Delta p - \left[ \left( \gamma_\parallel + \gamma_\perp - 2 \right) p_\perp + \left( \gamma_\parallel - 1 \right) \Delta p \right] \left[ \mathbf{b} \cdot \left( \mathbf{b} \cdot \nabla \right) \mathbf{v} \right] \\
 & + \eta \left[ \left( \gamma_\parallel + \gamma_\perp - 2 \right) \left( \mathbf{b} \cdot \mathbf{J} \right)^{2} - \left( \gamma_\perp - 1 \right) \left( \mathbf{J} \cdot \mathbf{J} \right) \right] + \frac{1}{3} \left( \gamma_\parallel - 2 \gamma_\perp + 1 \right) \nu \left| \nabla \times \mathbf{v} \right|^2 \nonumber
\end{align}

Linearizing the equation above, we immediately notice that the advection and viscous terms vanish, since our equilibrium field has $\mathbf{v}_0 = \mathbf{0}$. The second term on the right-hand side requires additional attention. The scalar products involving the magnetic field direction, $\mathbf{b}$, can be rewritten as
\begin{equation}
\mathbf{b} \cdot \left( \mathbf{b} \cdot \nabla \right) \mathbf{v} = \frac{1}{\left| \mathbf{B} \right|^2} \left[ \mathbf{B} \cdot \left( \mathbf{B} \cdot \nabla \right) \mathbf{v} \right].
\end{equation}

Applying the Taylor expansion of the inverse squared magnetic field magnitude, as shown previously, the resulting linearization of this term simplifies, largely due to the null equilibrium velocity field, to a simple form:
\begin{equation}
\mathbf{b} \cdot \left( \mathbf{b} \cdot \nabla \right) \mathbf{v} \approx \mathbf{B}_0 \cdot \left( \mathbf{B}_0 \cdot \nabla \right) \delta \mathbf{v}.
\end{equation}

The final part of the linearization of the equations is the ohmic heating. We have two terms there, namely, $\left( \mathbf{b} \cdot \mathbf{J} \right)^2$ and $\left(\mathbf{J} \cdot \mathbf{J}\right)$. Focusing on the first term, we can rewrite it as $\left( \mathbf{B} \cdot \mathbf{J} \right)^2 / \left| \mathbf{B} \right|^2$ and, using the approximation of $1 / \left| \mathbf{B} \right|^2$, obtain the linearized version of it,
\begin{equation}
\left( \mathbf{b} \cdot \mathbf{J} \right)^2 \approx 2 \left( \mathbf{B}_0 \cdot \mathbf{J}_0 \right) \left[ \left( \mathbf{B}_0 \cdot \delta \mathbf{J} \right) + \left( \mathbf{J}_0 \cdot \delta \mathbf{B} \right) - \left( \mathbf{B}_0 \cdot \mathbf{J}_0 \right) \left( \mathbf{B}_0 \cdot \delta \mathbf{B} \right) \right]
\end{equation}

The second term reduces to
\begin{equation}
\left(\mathbf{J} \cdot \mathbf{J}\right) \approx 2 \left( \mathbf{J}_0 \cdot \delta \mathbf{J} \right).
\end{equation}

Putting all together the linearized pressure difference equation results in
\begin{align}
\frac{ \partial \delta \Delta p}{\partial t} = & - \frac{1}{2} \left[ \left( \gamma_\parallel + \gamma_\perp - 2 \right) \beta_0 + \left( \gamma_\parallel - 1 \right) \Delta \beta_0 \right] \left[ \mathbf{B}_0 \cdot \left( \mathbf{B}_0 \cdot \nabla \right) \delta \mathbf{v} \right] \nonumber \\
& + 2 \eta \left\{ \left( \gamma_\parallel + \gamma_\perp - 2 \right) \left( \mathbf{B}_0 \cdot \mathbf{J}_0 \right) \left[ \left( \mathbf{B}_0 \cdot \delta \mathbf{J} \right) + \left( \mathbf{J}_0 \cdot \delta \mathbf{B} \right) - \left( \mathbf{B}_0 \cdot \mathbf{J}_0 \right) \left( \mathbf{B}_0 \cdot \delta \mathbf{B} \right) \right] - \left( \gamma_\perp - 1 \right) \left( \mathbf{J}_0 \cdot \delta \mathbf{J} \right) \right\} \nonumber
\end{align}

Using perturbations of the form $\delta f(t,\mathbf{r}) = \delta f(z) \exp[i k x + \sigma t]$, where $k$ and $\sigma$ denote the wavenumber along the x-direction and the growth rate of a given mode, respectively, and considering an equilibrium defined by the magnetic field profiles $B_0(z)$ and $B_g(z)$, zero equilibrium velocity, and uniform $\beta_0$ and $\Delta \beta_0$, after a relatively cumbersome derivation, we arrive at the final set of linearized nonideal gyrotropic-MHD equations. To improve clarity, we have marked the contributions from different processes, such as the pressure difference perturbation, the equilibrium pressure anisotropy, and the nonideal terms:

\begin{align}
& \sigma \delta v_y = \underbrace{i k B_0 \delta B_y + B_g' \delta B_z}_{\text{ideal MHD part}} \underbrace{- i k B_0 B_g \delta \Delta p}_{\text{pressure anisotropy fluctuations}} \underbrace{- \frac{\Delta \beta_0}{2} \left[ i k B_0 \left( 1 - 2 B_g^2 \right) \delta B_y + B_g' \delta B_z + 2 B_0^2 B_g \delta B_z' \right]}_{\text{equilibrium pressure anisotropy}} \nonumber \\
& \hspace{1.2in} \underbrace{+ \nu \left( \delta v_y'' - k^2 \delta v_y \right)}_{\text{viscous part}}
\end{align}
\begin{align}
& \sigma \left( \delta v_z'' - k^2 \delta v_z \right) =
 \underbrace{i k \left[ B_0 \left( \delta B_z'' - k^2 \delta B_z \right) - B_0'' \delta B_z \right]}_{\text{ideal MHD part}} \underbrace{- k^2 B_0 \left[ 2 B_0' \delta \Delta p + B_0 \delta \Delta p' \right]}_{\text{pressure anisotropy fluctuations}} \\
& \hspace{0.8in} \underbrace{- \frac{\Delta \beta_0}{2} \Bigl\{ i k \left[ B_0 \left( \delta B_z'' - k^2 \delta B_z \right) - B_0'' \delta B_z \right] - 2 \left[ i k B_0^2 \left[ B_0 \delta B_z'' + 3 B_0' \delta B_z' \right] \right.}_{\text{equilibrium pressure anisotropy}} \nonumber \\
& \hspace{0.4in} \quad \underbrace{\left. + k^2 B_0^2 B_g \delta B_y' + k^2 B_0 \left[ B_0 B_g' +2 B_0' B_g \right] \delta B_y \right] \Bigr\}}_{\text{equilibrium pressure anisotropy}} \underbrace{+ \nu \left( \delta v_{z}'''' - 2 k^2 \delta v_{z}'' + k^4 \delta v_{z} \right)}_{\text{viscous part}} \nonumber
\end{align}
\begin{align}
& \sigma \delta B_y = \underbrace{i k B_0 \delta v_y - B_g' \delta v_z}_{\text{ideal MHD part}} \underbrace{+ \eta \left( \delta B_y'' - k^2 \delta B_y \right)}_{\text{resistive part}}
\end{align}
\begin{align}
& \sigma \delta B_z = \underbrace{i k B_0 \delta v_z}_{\text{ideal MHD part}} \underbrace{+ \eta \left( \delta B_z'' - k^2 \delta B_z \right)}_{\text{resistive part}}
\end{align}
\begin{align}
& \sigma k \delta \Delta p = \underbrace{\frac{1}{2} \left( \gamma_\parallel + \gamma_\perp - 2 \right) \beta_0 k B_0 \left( B_0 \delta v_{z}' - i k B_{g} \delta v_{y} \right)}_{\text{plasma-$\beta$ dependence}} \underbrace{+ \frac{1}{2} \left( \gamma_\parallel - 1 \right) \Delta \beta_0 k B_0 \left( B_0 \delta v_{z}' - i k B_{g} \delta v_{y} \right)}_{\text{equilibrium pressure anisotropy}} \label{eq:dp_lin} \\
& \hspace{0.1in} \underbrace{- 2 \eta \left\{ \left( \gamma_\parallel + \gamma_\perp - 2 \right) \left( B_g B_0' - B_0 B_g' \right) \left[ k \left( B_{0}' \delta B_{y} - B_0 \delta B_{y}' \right) \right. + i \left( B_g \left( \delta B_{z}'' - k^2 \delta B_{z} \right) - B_{g}' \delta B_{z}' \right) \right.}_{\text{ohmic heating}} \nonumber \\
&  \hspace{0.24in} \underbrace{\left. - \left( B_g B_0' - B_0 B_g' \right) \left( i B_0 \delta B_{z}' + k B_g \delta B_{y} \right) \right] \left. + \left( \gamma_\perp - 1 \right) \left[ k B_{g}' \delta B_{y}' + i B_{0}' \left( \delta B_{z}'' - k^{2} \delta B_{z} \right) \right] \right\}}_{\text{ohmic heating}} \nonumber
\end{align}

In the above equations, $'$, $''$, and $''''$ denote the first-, second-, and fourth-order derivatives with respect to $z$, respectively.


\bibliography{ms}{}
\bibliographystyle{aasjournal}



\end{document}